\documentclass[pra,twocolumn]{revtex4-1}

\usepackage{graphicx}
\usepackage{amsmath}
\usepackage{color}

\begin{document}

\title{Coherent adiabatic transport of atoms in radio-frequency traps}

\author{T.~Morgan, B.~O'Sullivan, and Th.~Busch}

\affiliation{Department of Physics, University College Cork,
  Cork, Ireland}

\begin{abstract}
  Coherent transport by adiabatic passage has recently been suggested
  as a high-fidelity technique to engineer the centre-of-mass state of
  single atoms in inhomogenous environments. While the basic theory
  behind this process is well understood, several conceptual
  challenges for its experimental observation have still to be
  addressed. One of these is the difficulty that currently available
  optical or magnetic micro-trap systems have in adjusting the tunneling
  rate time-dependently while keeping resonance between the asymptotic
  trapping states at all times. Here we suggest that both requirements
  can be fulfilled to a very high degree in an experimentally
  realistic setup based on radio frequency traps on atom chips. We
  show that operations with close to 100\% fidelity can be achieved
  and that these systems also allow significant improvements for
  performing adiabatic passage with interacting atomic clouds.
\end{abstract}



\maketitle

\section{Introduction}
\label{sec:Introduction}

Going beyond nano-technologies and engineering quantum systems on the
basis of single particles has in recent years been one of the most
exciting and active areas of physics \cite{Nielsen:00}. Due to the
fragile nature of single particle quantum states, quantum engineering
techniques need to be fault tolerant and lead to high fidelities on
every application to avoid the large and costly overhead that comes
with error correction schemes \cite{Steane:96}. One class of
techniques that can achieve this are so-called adiabatic techniques
and their use in optical systems has been widely investigated in the
past. In particular, Stimulated Raman Adiabatic Passage (STIRAP) is
one adiabatic technique that allows to transfer the populations from
one electronic state to another with 100\% fidelity
\cite{Bergmann:98}. It relies on the existence of a so-called dark
state in a three level system and requires a counter-intuitive pulse
sequence to coherently couple the individual levels.

Recently, it has been shown that similar techniques can, in principle,
be used to control the quantised centre-of-mass state of single
particles \cite{Eckert:04,Greentree:04,Eckert:06}. This atom-optical
analogue has been dubbed Coherent Transport by Adiabatic Passage
(CTAP) and while the possibility of observing this process has
received significant attention \cite{Rab:08,OSullivan:10}, the
conditions that have to be fulfilled for its observation are currently
hard to achieve experimentally. In particular, all states involved are
required to be in resonance during the whole process. However, since
the strength of the tunnel-coupling is usually adjusted by changing
the distance between the microtraps, which leads to significant
overlap of the neighbouring trapping potentials, the eigenstates
become time-dependent. Several solutions to the problem have been
suggested, with all involving significant experimental resources or
restrictions on the parameter space
\cite{Eckert:04,Eckert:06,Rab:08}. A similar process coupling
classical light between optical waveguides has recently been
experimentally demonstrated \cite{Longhi:06,Longhi:07,Rangelov:09}.

Here we propose a simple experimental setup that fulfills all
necessary conditions to observe CTAP for cold atoms.  Our proposal is
based on radio frequency (rf) traps, which have recently become one of
the most versatile tools for trapping cold atoms
\cite{Zobay:01,Schumm:05}. The advantage of rf-systems is that their
physics is well known, they are relatively benign systems to work with
experimentally and are widely available today. They not only allow to
create standard trapping potentials \cite{Zobay:01}, but can also be
used to coherently manipulate matter waves
\cite{Schumm:05,Hofferberth:07} or create complicated, non-standard
trapping geometries \cite{Courteille:06,Fernholz:07,Lesanovsky:06}.

We will also show that our setup offers the possibility for extending
adiabatic techniques to clouds of interacting atoms. The presence of
interaction between the atoms introduces non-linearities into the
system \cite{Liu:06} which have been shown to inhibit the
effectiveness of CTAP in transporting atoms \cite{Rab:08}. Several
strategies to adjust the process and to allow transport in the
presence of these non-linear interactions have been suggested, for
example a fixed detuning between the potential wells
\cite{Graefe:06}. Here we will show that dynamically controlling the
detuning between the potentials provides a marked improvement in the
state transfer efficiency over both regular and fixed detuning CTAP.

In the following we will first briefly review the idea of CTAP for
ultracold atoms. In section \ref{sec:RFTraps} we will outline the
theoretical description of rf-trapping and describe the system needed
for CTAP. In section \ref{sec:AtomicTransport} we demonstrate atomic
transport in this system and show that the process allows high
fidelity atomic transport in contrast to the intuitive method, which
fails. In section \ref{sec:Nonlinear} we examine the transport
of an interacting atomic cloud and how the presence of non-linear
interaction can be compensated for by dynamic detuning. Finally we
conclude.

\section{Coherent Transport Adiabatic Passage}
\label{sec:CTAP}

\begin{figure}
  \includegraphics[width=\linewidth]{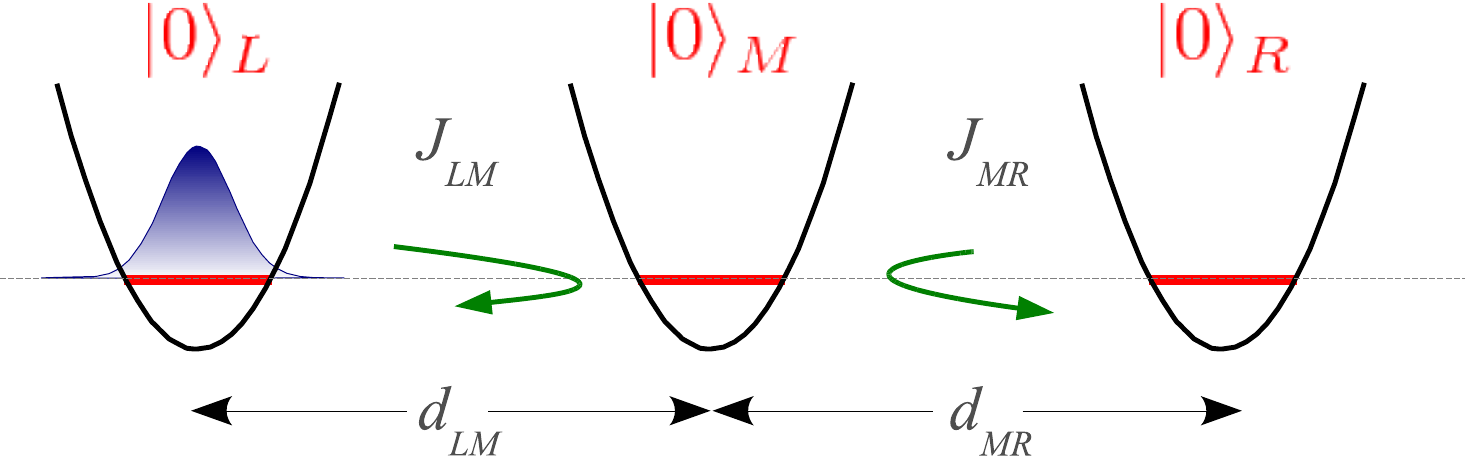}
  \caption{Schematic of the CTAP process for an atom in the left trap. 
  Reducing the distances between the traps leads to an increase in the 
  tunnelling strengths.}
  \label{fig:Schematic}
\end{figure}

To briefly review the process of adiabatic population transfer let us
consider a system of three ground states in three identical
microtraps, $|0\rangle_L, |0\rangle_M$ and $|0\rangle_R$ (see
Fig.~\ref{fig:Schematic}). In a linear arrangement the only tunnel
couplings that are significant are $J_{LM}$ for the transition $\vert
0 \rangle_L$ $\to$ $\vert 0 \rangle_M$ and $J_{MR}$ for $\vert 0
\rangle_M$ $\to$ $\vert 0 \rangle_R$. Assuming that the three states
are in resonance when isolated, the Hamiltonian for such a system is
given by
\begin{equation}
  \label{eq:wgHamiltonian}
  H(t)=\hbar
    \left(\begin{array}{ccc}
    0             & -J_{LM}(t) & 0  \\
    -J_{LM}(t)  & 0            & -J_{MR}(t) \\
    0             & -J_{MR}(t) & 0  
  \end{array}\right) \;.
\end{equation}
and a smooth time dependence of the tunnelling coupling pulses can be
achieved by continuously changing the distances between the traps,
$d_{LM}(t)$ and $d_{MR}(t)$. The eigenstates of this Hamiltonian are
very well known \cite{Bergmann:98} and of particular interest to our
work here is the so-called dark state
\begin{equation}
  |d\rangle=\cos\theta|0\rangle_L-\sin\theta|0\rangle_R\,
\end{equation}
for which the  mixing angle $\theta$ is given by
\begin{equation}
  \tan\theta=J_{LM}/J_{MR}.
\end{equation}
This state has a non-degenerate zero eigenvalue and therefore an
adiabatic evolution will guarantee that the system, once prepared in
$|d\rangle$, will always stay in it. Note that as the only
contribution to $|d \rangle$ from the state $|0\rangle_M$ comes
through the mixing angle, the system has zero probability to be found
in $|0\rangle_M$ at any time.

The CTAP process can now be understood by considering an atom
initially in the state $|0\rangle_L$. Increasing and decreasing
$J_{MR}$ before $J_{LM}$, which is counter-intuitive to traditional
tunneling schemes, continuously decreases the population in state
$|0\rangle_L$ and increase the population in state $|0\rangle_R$,
leading to a 100\% transfer at the end of the process.

It is worth to stress again the conditions that have to be fulfilled
for the above dynamics to occur. Firstly, the process must be
adiabatic with respect to the energy level splitting in the harmonic
oscillators, which means that the movement of the traps has to be slow
and the whole process must take longer than $\omega^{-1}_\text{HO}$,
where $\omega_\text{HO}$ is the harmonic oscillator frequency of the
individual traps. As typical numbers of $\omega_\text{HO}$ for
microtraps are in the kHz regime, this means that the time required
for this process is much shorter than lifetimes of the trapped atoms,
which makes this process a promising tool for quantum information. The
other condition we require, as previously mentioned, is that all
single trap states are in resonance at any point in time, which is
difficult to achieve once the trapping potentials start to overlap.

In the next section we will demonstrate how the second condition can
be fulfilled in an experimentally realistic system using radio
frequency potentials.

\section{\label{sec:RFTraps} Radiofrequency Trapping}

Radio frequency trapping relies on the process of coupling magnetic
sublevels in the presence of an inhomogeneous magnetic field
\cite{Zobay:01,Courteille:06,Hofferberth:07,Schumm:05}. Consider a
hyperfine atomic groundstate with total spin $F=\frac{1}{2}$. In the
presence of the magnetic field the two hyperfine sublevels
$m_F=\frac{1}{2}$ and $m_F'=-\frac{1}{2}$ are split by an amount
$\mu_B g_F m_F B$, where $g_F$ is the atomic g-factor of the hyperfine
level and $\mu_B$ is the Bohr magneton. Irradiating such a system with
a linearly polarized radio frequency, ${\bf B}_\text{rf} \cos(\omega
t)$, couples the sublevels $|\frac{1}{2},\frac{1}{2}\rangle
\leftrightarrow |\frac{1}{2},-\frac{1}{2}\rangle$ with spatial
resolution due to the spatial dependence of the magnetic field. Here
we will concentrate on a one dimensional description of such a
process, which is valid when the radio frequency and magnetic field to
be orthogonal to each other.  Assuming the inhomogeneous magnetic
field to be oriented in the $x$-direction, $B$=$B(x)$, the Hamiltonian
of the coupled system can be written as
\begin{equation}
  \label{eq:rfHamiltonian}
  H(x)=\frac{1}{2}
    \left(\begin{array}{cccc}
    \mu_B g_F B(x) - \hbar \omega & 
    \hbar \Omega  \\
    \hbar \Omega & 
    -\mu_B g_F B(x) + \hbar \omega  
          \end{array}\right) \;,
\end{equation}
where the strength of the coupling is given by the Rabi frequency
\cite{Ketterle:96}
\begin{equation}
  \label{eq:Rabi Frequency}
  \Omega=\frac{\mu_B g_F}{4\hbar}| {\bf B}_\text{rf} \times 
         \hat{e}_B | \sqrt{F(F+1) - m_F m^{'}_F}\;,
\end{equation}
and where $\hat{e}_B$ is the orientation of the local static magnetic
field. The eigenvalues of this Hamiltonian are \cite{Courteille:06}
\begin{align}
  E_{\pm}(x) &= \pm \frac{1}{2} \sqrt{ \hbar^2 \Omega^2 
              + \lbrack \mu_B g_F B(x) - \hbar \omega \rbrack^2}\;,\\
            &\approx \pm \frac{1}{2} \lbrack \mu_B g_F B(x) 
                   - \hbar \omega \rbrack 
                   \pm \frac{\hbar^2 \Omega^2}
                   {4 \lbrack \mu_B g_F B(x) - \hbar \omega \rbrack}\;,
\end{align}
where the second expression is valid far from the resonance, $\hbar
\Omega \ll \lbrack \mu_B g_f B(x) - \hbar \omega \rbrack$. The second 
term in the expression can be viewed as a Stark shift on the energy levels.

\begin{figure}[tb]
  \includegraphics[width=\linewidth]{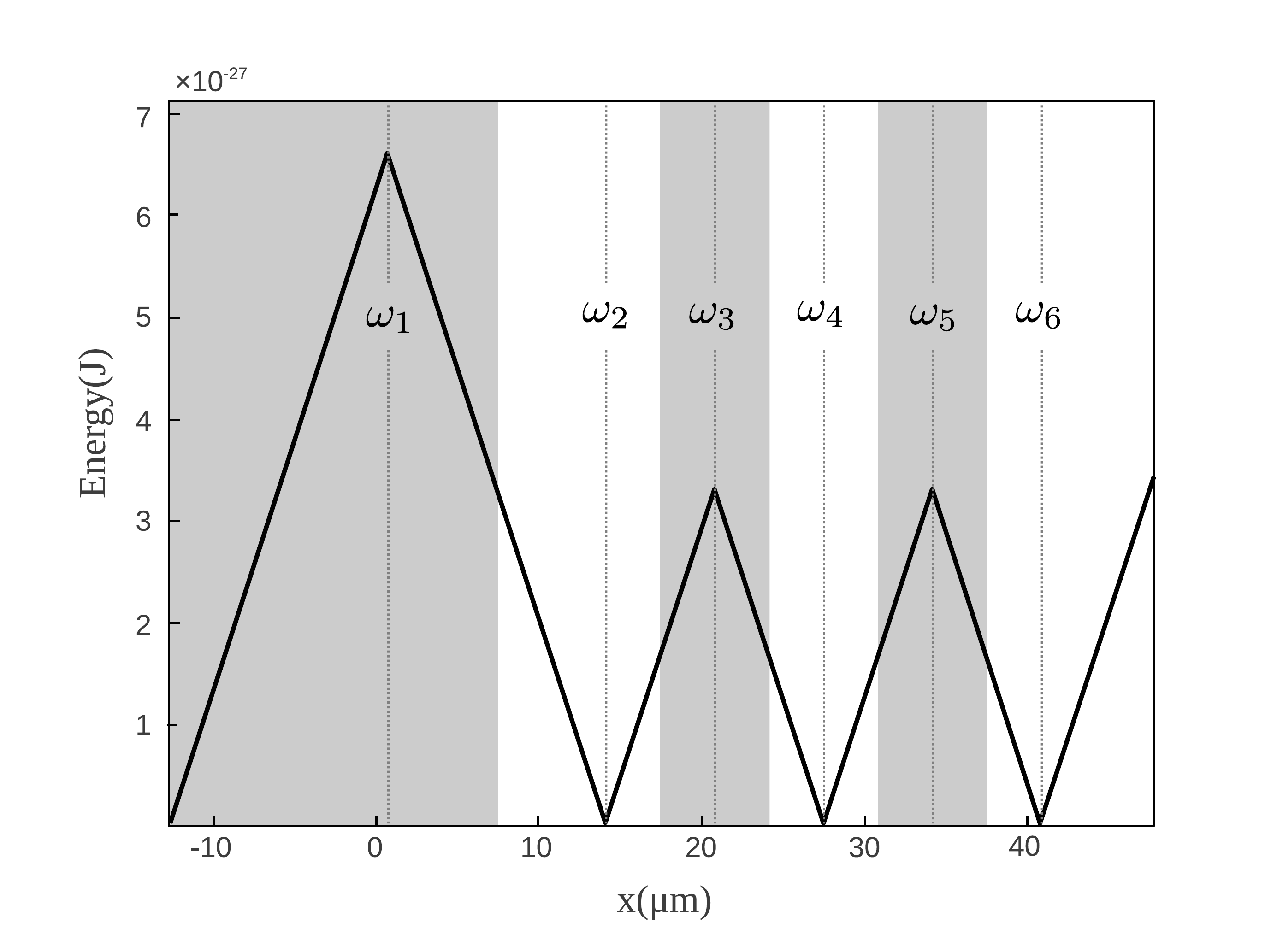}
  \caption{Trapping potential created by six radio-frequencies
    $\omega_1=1000$ kHz and $\omega_n=2\pi n\times 10000$ kHz,
    $n=2:6$. Their resonance-positions are marked by the broken
    vertical lines and the range over which they are applied is
    indicated by the grey and white zones. The magnetic field gradient
    has the strength $b=213$ Gcm$^{-1}$ and $g_F=-\frac{1}{2}$ for the
    $^{87}$Rb ground state $^{2}S_\frac{1}{2}$. The Rabi frequency is
    chosen to be $2 \pi \times 50$ kHz. The traps resemble harmonic
    oscillator potentials close to each minima. }
\label{fig:rfSchematic}
\end{figure}

To create a multi-well potential it is necessary to use several
frequencies and the above analysis will become significantly more
complicated. However, if we assume that the individual frequencies are
spaced sufficiently far apart and have low Rabi frequencies with
respect to the detuning, we can approximate the dynamics locally by
considering only the nearest resonance frequency,
$\omega(x)=\omega_{n(x)}$ \cite{Courteille:06}. Formally this means
that $n$ is chosen such that $\lbrack \mu_B g_F B(x) - \hbar
\omega_{n(x)} \rbrack$ is minimized at any position $x$. The effects
of the combined Stark shifts, produced by the frequencies not closest
to resonance, can then be summed up as \cite{Courteille:06}
\begin{equation}
  L_n(x)=\sum_{j\not=n} \frac{\hbar^2 \Omega^2}
         {4 \lbrack \mu_B g_F B(x) - \hbar \omega_{j(x)} \rbrack},
\end{equation}
so that the eigenvalues are given by
\begin{equation}
  E_{\pm}(x) =\pm\frac{1}{2}\sqrt{ \hbar^2 \Omega^2 + \lbrack \mu_B g_F B(x) 
              - \hbar \omega +2 L_n(x) \rbrack^2}.
\end{equation}
From this, and considering the that the couplings are strong enough to
yield Landau-Zener transition probability close to unity, the
resulting adiabatic potential is given by
\begin{equation}
  V_{ad,\pm}(x)=(-1)^{n(x)} \left[ E_\pm(x) \mp \frac{\hbar \omega_{n(x)}}{2} \right] 
               \mp \sum^{n(x)-1}_{k=1}(-1)^k \hbar \omega_k.
\end{equation}
To produce a radio frequency potential with three minima along the
$x$-direction we will need six different radio frequencies. In the
following we will assume that the 1D linear magnetic field is given by
$B(x)=bx$ where $b$ is the magnetic field gradient. For convenience we
choose five of the six radio frequencies to be equally spaced
initially, $\omega_n=2n\pi \times 10000$ kHz ($n=2:6$), which produces
three equidistant minima.  The value of the first radio frequency
$\omega_1$ does need to have the same distance as the other
frequencies, as its value only controls the height of the first
maximum (see Fig.~\ref{fig:rfSchematic}) and can therefore be adjusted
without changing the trap geometry in the area where tunneling takes
place. For our potential we set $\omega_1=1000$ kHz and in
Fig.~\ref{fig:rfSchematic} we indicate the {\sl local} frequencies and
show the resulting adiabatic potential in the positive $x$-direction.

\section{Adiabatic Passage}
\label{sec:AtomicTransport}

In this section we will apply the CTAP procedure to a single atom
trapped in a three well rf-potential. We will show that the strong
decay of the influence of the radio frequencies away from their
respective resonance points allows us to fulfill the resonance
condition between the asymptotic eigenstates at all times during the
process. While the Stark shift from neighbouring resonances cannot be
neglected, it is small enough to not destroy the process.

Movement of the traps is achieved by changing the individual radio
frequencies that are associated with each trap. Traditionally for CTAP
the middle trap is chosen to be at rest and the two outer ones are
moving in and out (see also Fig.~\ref{fig:Schematic}). Here we will
choose a slightly different, but of course completely analogous, route
in that we keep the position of the left trap fixed. This allows us to 
keep the values of the minima equal which is essential to satisfy the 
condition of resonance between all traps.

\begin{figure}[tb]
  \includegraphics[width=\linewidth]{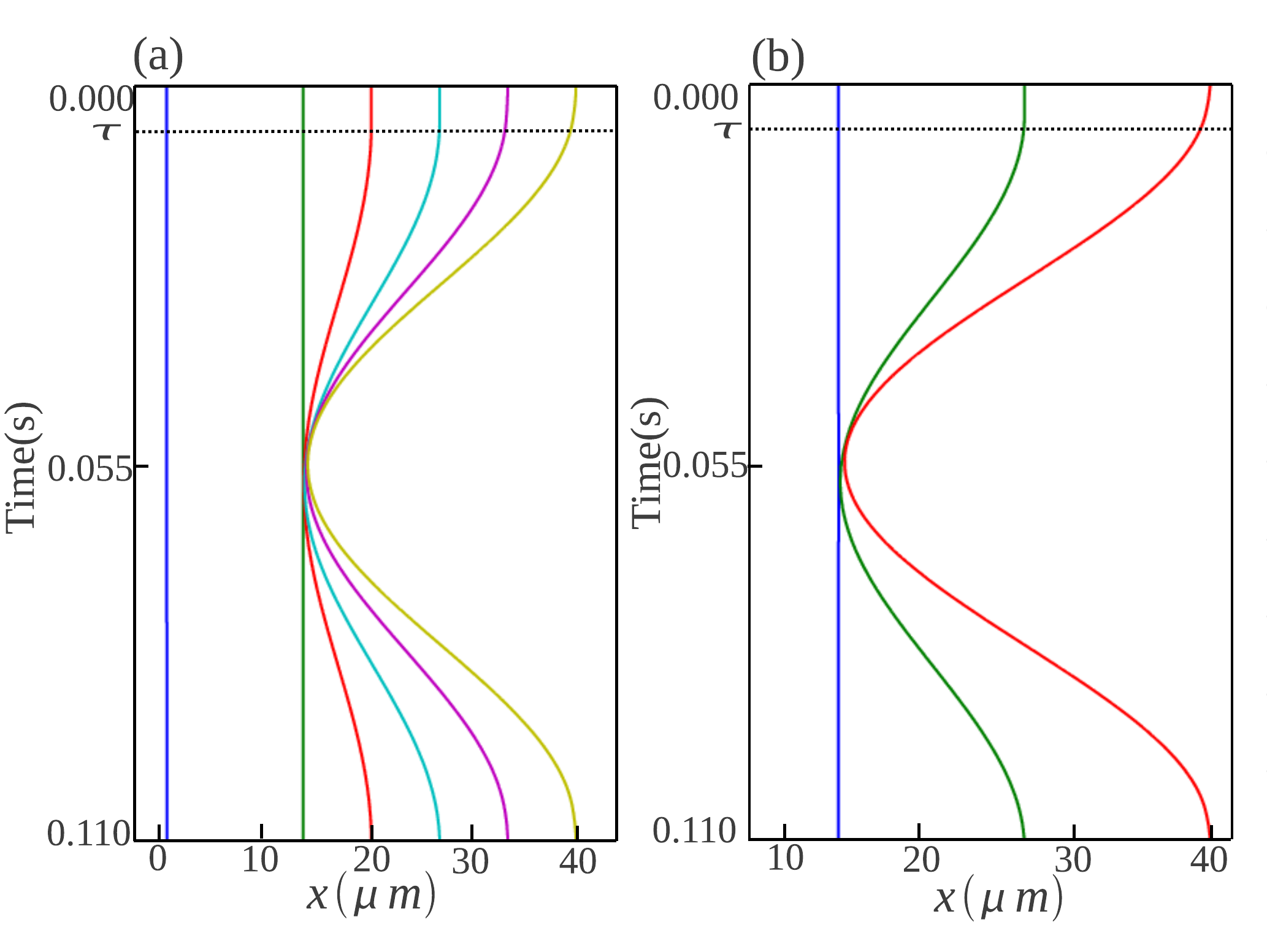}
  \caption{(a) Radiofrequencies, $\omega_n$, as a function of time to
    achieve the counter-intuitive positioning sequence. (b) Positions
    of the trap minima as a function of time. The left trap remains
    stationary while the other two traps move towards it. The delay in
    the movement of middle trap in comparison to the right trap
    ($\tau=0.0055$ s) is indicated by the broken line.}
\label{fig:FreqAndMins}
\end{figure}

In order to achieve CTAP when moving the traps in this non traditional
manner the approach of the right trap towards the middle must start
earlier than that of the approach of the middle trap to the left. One
therefore initially only changes the frequencies $\omega_5$ and
$\omega_6$, which determine the shape and position of the right hand
side trap. After a delay $\tau$, the two frequencies $\omega_3$ and
$\omega_4$ are changed as well, allowing to move the middle trap
towards the left. Due to the adiabatic nature of the process the exact
shape of this time-dependent frequency adjustment, $f(t)$, does not
matter and we can formalise this process as
\begin{subequations}
  \begin{align}
    \omega_1(t)&=\omega_1(t_0),\\
    \omega_2(t)&=\omega_2(t_0),\\
    \omega_3(t)&=\omega_3(t_0)-\frac{1}{2}f(t-\tau)\theta(t-\tau),\\
    \omega_4(t)&=\omega_4(t_0)-f(t-\tau)\theta(t-\tau),\\ 
    \omega_5(t)&=\omega_5(t_0)-\frac{1}{2}f(t)-f(t-\tau)\theta(t-\tau),\\ 
    \omega_6(t)&=\omega_6(t_0)-f(t) - f(t-\tau)\theta(t-\tau)\,
  \end{align}
\label{eq:RFAdjustmentLinear}
\end{subequations}
where $\theta(t)$ is the Heaviside step function. In
Fig.~\ref{fig:FreqAndMins}(a) these changes are shown for the typical
system considered here and the resulting movements of the trap minima
are displayed in Fig.~\ref{fig:FreqAndMins}(b). As can been seen, the
minimum of the left trap remains stationary while the other traps are
moving towards and away from it. The resulting movement between
neighbouring traps exactly fulfills the requirement of the CTAP
process, leading to the desired increase and decrease in the tunnelling strength
between initially the middle and right traps before the increase and decrease in
tunnelling strength between the left and middle traps.

To demonstrate adiabatic passage for single atoms and for typical
experimentally realistic parameters, we will in the following show the
results of numerical simulations of the full Schr\"odinger
equation. We choose a single $^{87}$Rb atom to be initially located in
the centre-of-mass ground state of the left trap and start the process
described in eq.~\eqref{eq:RFAdjustmentLinear} with an initial
separation between the radio frequencies of 10000 kHz. The minimum
distance to which the frequencies approach each other is 200 kHz,
which ensures that we are always in the regime of tunnelling
interaction, as the minimum barrier height between the individual
traps is $5.3313 \times 10^{-29}$~J at the point of closest approach,
compared to the ground state energy of $1.3615 \times 10^{-29}$~J. The
form of the adjustment function $f(t)$ is taken to be a cosine and for
numerical simplicity we restrict ourselves to ones spatial dimension.

\begin{figure}[tb]
  \includegraphics[width=\linewidth]{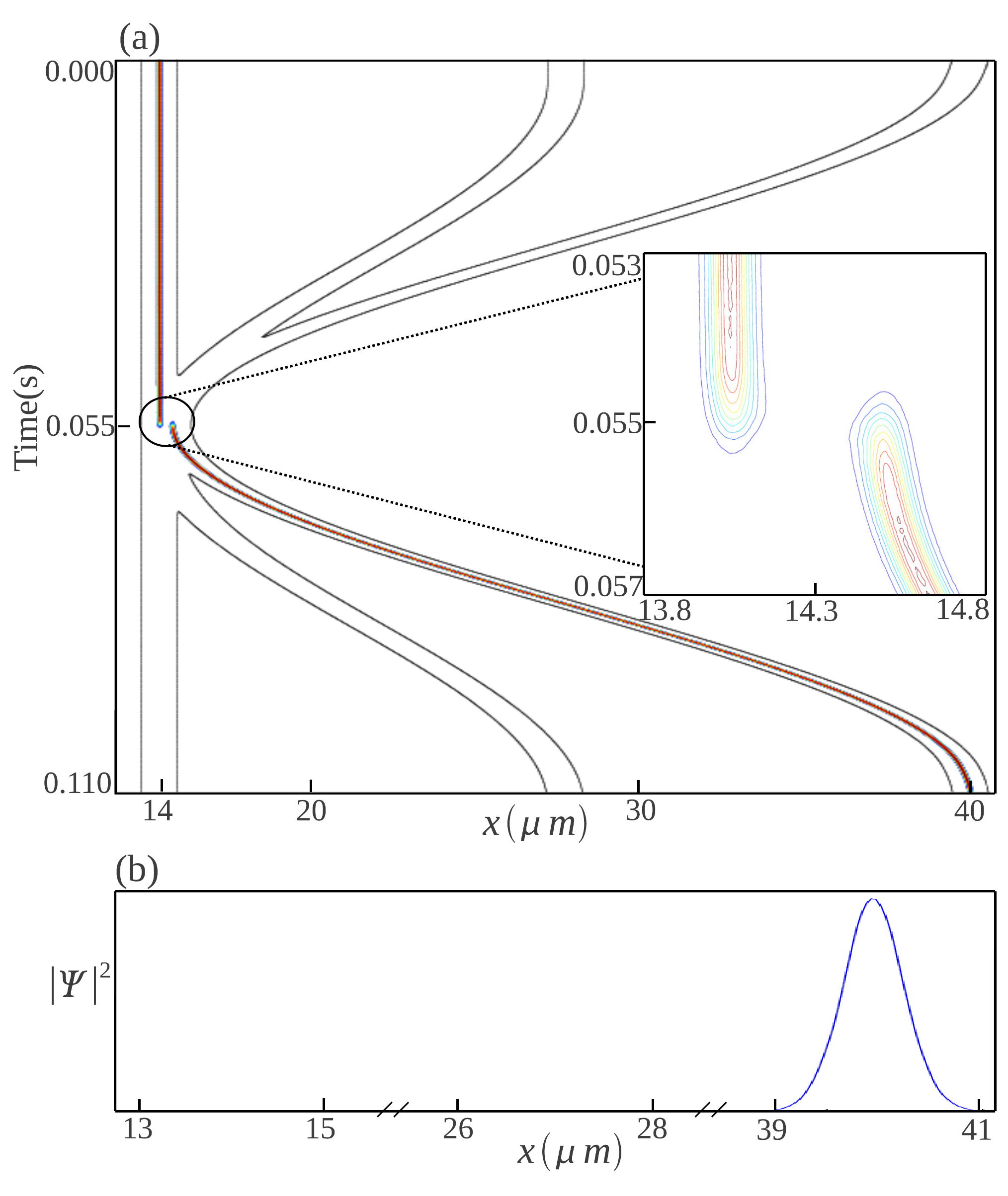}
  \caption{(a) Probability density for a single atom initally located
    in the trap on the left hand side with respect to time for {\sl
      counter-intuitive} trap movement. The inset shows the tunnelling
    area in greater detail. (b) Density of the final state in each of
    the three traps.}
\label{fig:CI1Atom}
\end{figure}

In Fig.~\ref{fig:CI1Atom} we show the probability density function
with respect to time for the CTAP process. The overall time for this
process is chosen to be $T=0.11s$ which is large compared to the
approximate harmonic oscillator frequency of the individual traps of
$\omega^{-1}_\text{HO}\approx 4 \times 10^{-6}$~s, and we are
therefore assured to be at all times in the dark eigenstate of the
system. This can also be seen from the fact that the probability for
being in the middle trap at any time is zero. The process leads to
high fidelity population transfer and an absence of Rabi oscillations.

\begin{figure}[tb]
 \includegraphics[width=\linewidth]{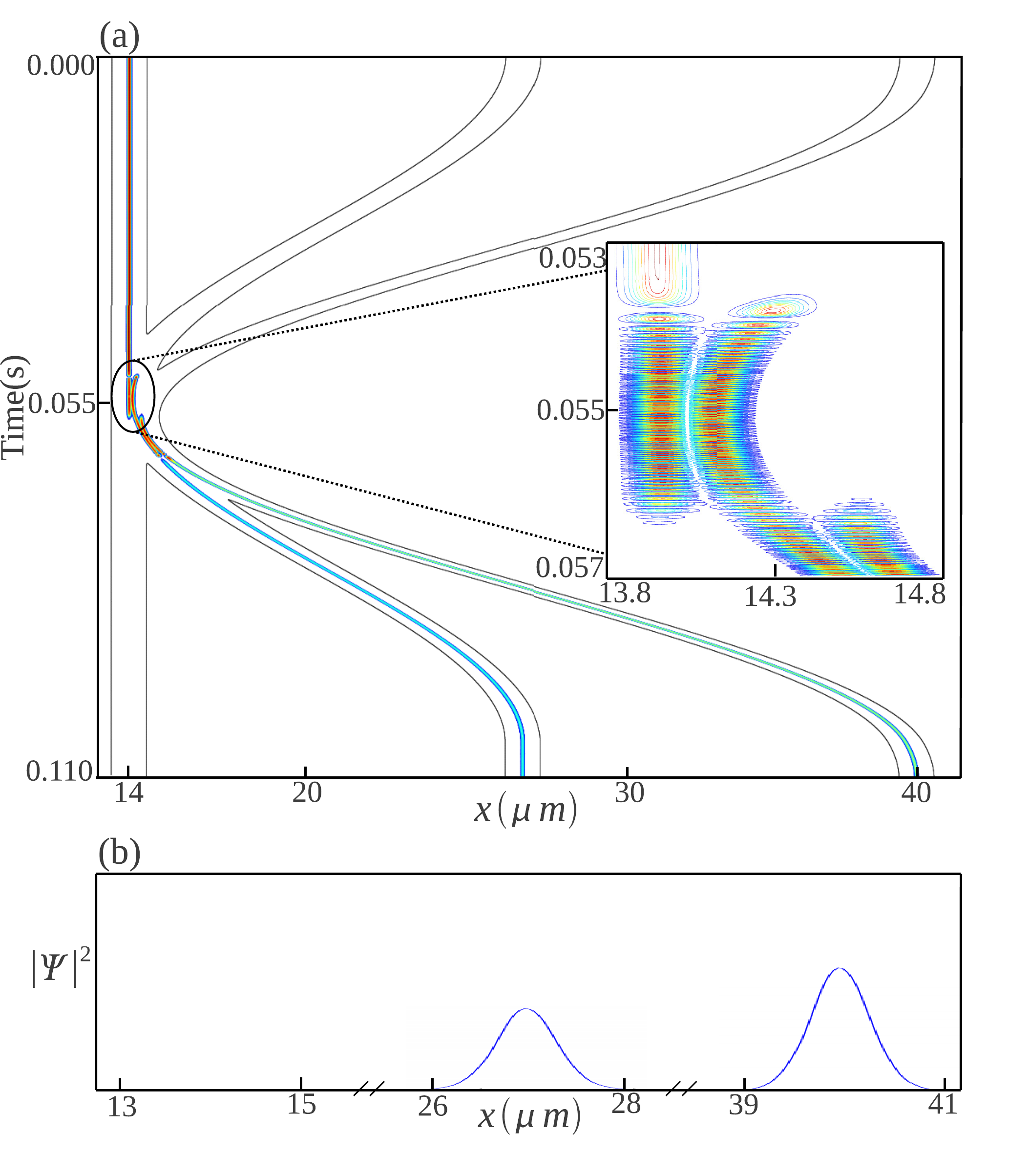}
 \caption{(a) Probability density for a single atom initally located
   in the trap on the left hand side with respect to time for {\sl
     intuitive} trap movement. The inset shows the tunnelling area in
   greater detail, where Rabi oscillations between neighbouring traps
   are clearly visible. (b) Density of the final state in each of the
   three traps.}
\label{fig:I1Atom}
\end{figure}

To compare the above situation to a process in which direct tunnelling
between two neighbouring traps plays an important role, we show in
Fig.~\ref{fig:I1Atom} the results of the same process, this time
however using an intuitive trap-movement. The direct tunnelling is
clearly manifest in the appearance of Rabi oscillations between the
traps and the process therefore does not deliver the required robust
population transfer. In fact, the final state becomes highly
susceptible to variations of the system parameters \cite{Mompart:03}.

We have confirmed that these results are representative for a large
range of parameters, making rf-traps ideal systems to investigate
adiabatic processes in all generality.

\section{Non-linear Systems}
\label{sec:Nonlinear}

The extension of adiabatic methods to non-linear systems is of large
importance not only to describe experimental situations, but also for
the understanding of the underlying physical principles
\cite{Liu:06,Graefe:06,Smerzi:97,Marino:99}.  In this section we show
how CTAP can be used with time-dependent potentials to coherently
transport a cloud of interacting, Bose-condensed atoms. For this, we
treat the adiabatic process as a series of stationary states which can
be described by the time-independent Gross-Pitaevskii equation
\begin{equation}
  \label{eq:GP}
  \mu\Psi(x)=\left(-\frac{\hbar^2}{2m}\nabla^2 
                          + V(x) + g_{1D}\vert\Psi\vert^2\right)\Psi(x)\;,
\end{equation}
where $V(x)$ is the external and $\mu$ the chemical potential at each
respective point in time. The one-dimensional interaction strength
between bosons with a three-dimensional s-wave scattering length $a_s$
is given by and $g_{1D}=\frac{4 N\hbar^2 a_s}{m a_\perp}(a_\perp-C
a_s)^{-1}$ \cite{Olshanii:98}. The trap width in the radial direction
is given by $a_\perp$ and $C\approx1.4603$. In the three level
approximation the Hamiltonian can therefore be written as
\begin{equation}
  \label{eq:NLHamiltonian}
  H(t)=\hbar
  \left(\begin{array}{ccc}
      \hbar\omega_L+\mu_L   & -J_{LM}(t)    & 0  \\
      -J_{LM}(t)            & \hbar\omega_M & -J_{MR}(t) \\
      0                     & -J_{MR}(t)    & \hbar\omega_R+\mu_R  
    \end{array}\right) \;,
\end{equation}
where $\mu_{L,R}$ are the chemical potentials associated with the
atomic clouds in left or right trap and $\omega_{L,M,R}$ are the
harmonic oscillator frequencies associated with the individual traps.
Because the particle number in each individual trap is a function of
time, the associated change in the chemical potentials will detune the
resonance between the energies. To be able to compensate for this we
will in the following allow for the trapping frequencies to be
functions of time as well.  Starting with a cloud of atoms in the left
trap and then attempting to perform CTAP, it is clear that the
chemical potential $\mu_L$ will decrease over time, while $\mu_R$ will
increase. As at all times the uncoupled traps have to be in resonance,
one can see that a time-dependent adjustment of the trapping
frequencies $\omega_{L,R}$ can allow us to compensate for this
change. However, in order to be able to make the three state
approximation, we need to make sure that $\mu<\hbar\omega_{L,M,R}$ at
all times, for all values of $\mu$ and $\omega_{L,M,R}$. This means in
practice that the process is limited to cold atomic clouds with small
non-linearities.

\begin{figure}[tb]
  \includegraphics[width=\linewidth]{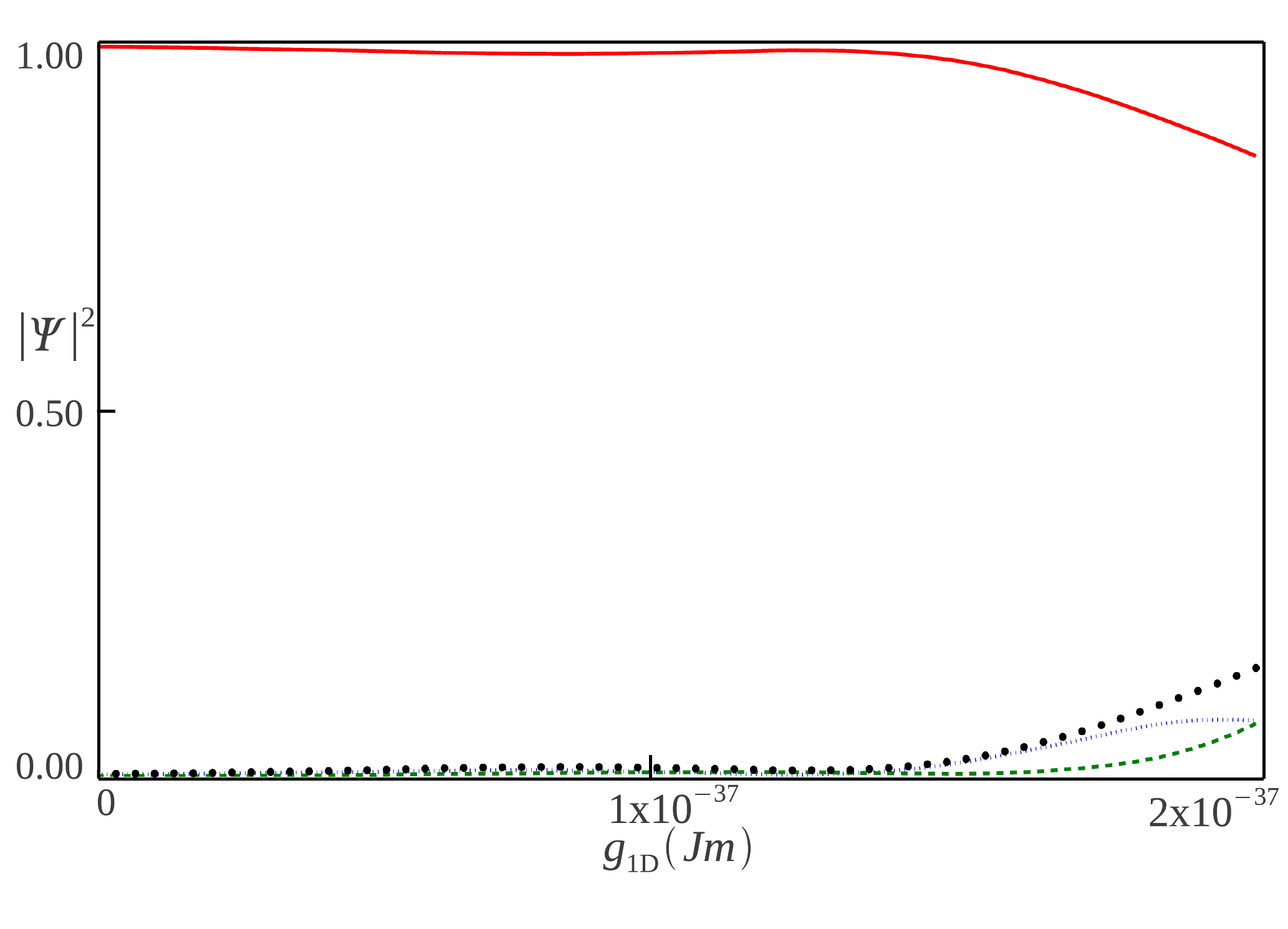}
  \caption{Final population in left (verticle dashed line, blue),
    middle (horizontal dashed line, green) and right (solid line, red)
    traps with increasing interaction strength. The dotted black line
    shows the total population not occupying the target (right)
    trap. The maximum value of $g_{1D}$ corresponds to $\mu=1.4318
    \times 10^{-29}$ J which is smaller than $\hbar\omega_{L,M,R}$ at
    all times.}
  \label{fig:NAFinalPopulationNL}
\end{figure}

Using the same radio frequency potential as in the linear case, we
place a cloud of interacting $^{87}$Rb atoms in the ground state of
the left trap by determining the solution to the Gross-Pitaevskii
equation for an isolated trapping potential. To show the influence of
the non-linear behaviour, we first carry out the same
counter-intuitive trap movements as in the linear section {\it
  without} time-dependent change in the trapping frequencies. In
Fig.~\ref{fig:NAFinalPopulationNL} we show the final populations in
the individual traps as a function of increasing values for
$g_{1D}$. It is immediately obvious that even for weak interactions
the non-linear term is disruptive to the process of CTAP. In fact, for
$g_{1D}=2 \times 10^{-37}$ Jm the state transfer efficiency is reduced
to $84\%$. Choosing a typical radial trap width of $~130$ nm, this
value of $g_{1D}$ corresponds to $N=2$ for $^{87}$Rb. 

As we have already stated above, a possible strategy to combat this
effect of changing system energies in the individual traps is to
adjust the individual trapping frequencies.  To restore resonance for
a changing chemical potential one can adjust the trapping frequency to
make sure that at any point in time $\hbar\omega_L(t)+\mu_L(t) =
\hbar\omega_M = \hbar\omega_R(t)+\mu_R(t)$. However, determining the required
adjustments is not a simple exercise for at least two reasons. First, the 
density dependence of the chemical potential will
prevent this change from simply being linear in time and, secondly, a
conceptual difficulty in determining the individual chemical
potentials arises when the traps are close together.  While one could
try to calculate the chemical potential, and therefore the on-site
energies, in all traps at all times to a good approximation, this is certaintly
not experimentally possible. In the following, we therefore
suggest a simple functional form for dynamically detuning the outer
traps and we show that it allows us to achieve significantly higher
transfer than possible without adjustments. A similar idea, however
without time-dependence, was recently proposed by Graefe {\sl et al.}
\cite{Graefe:06}, who showed that by detuning the left and the right
trap by the same fixed amount throughout the process an improved
transfer of population can be achieved.

\begin{figure}
  \includegraphics[width=\linewidth]{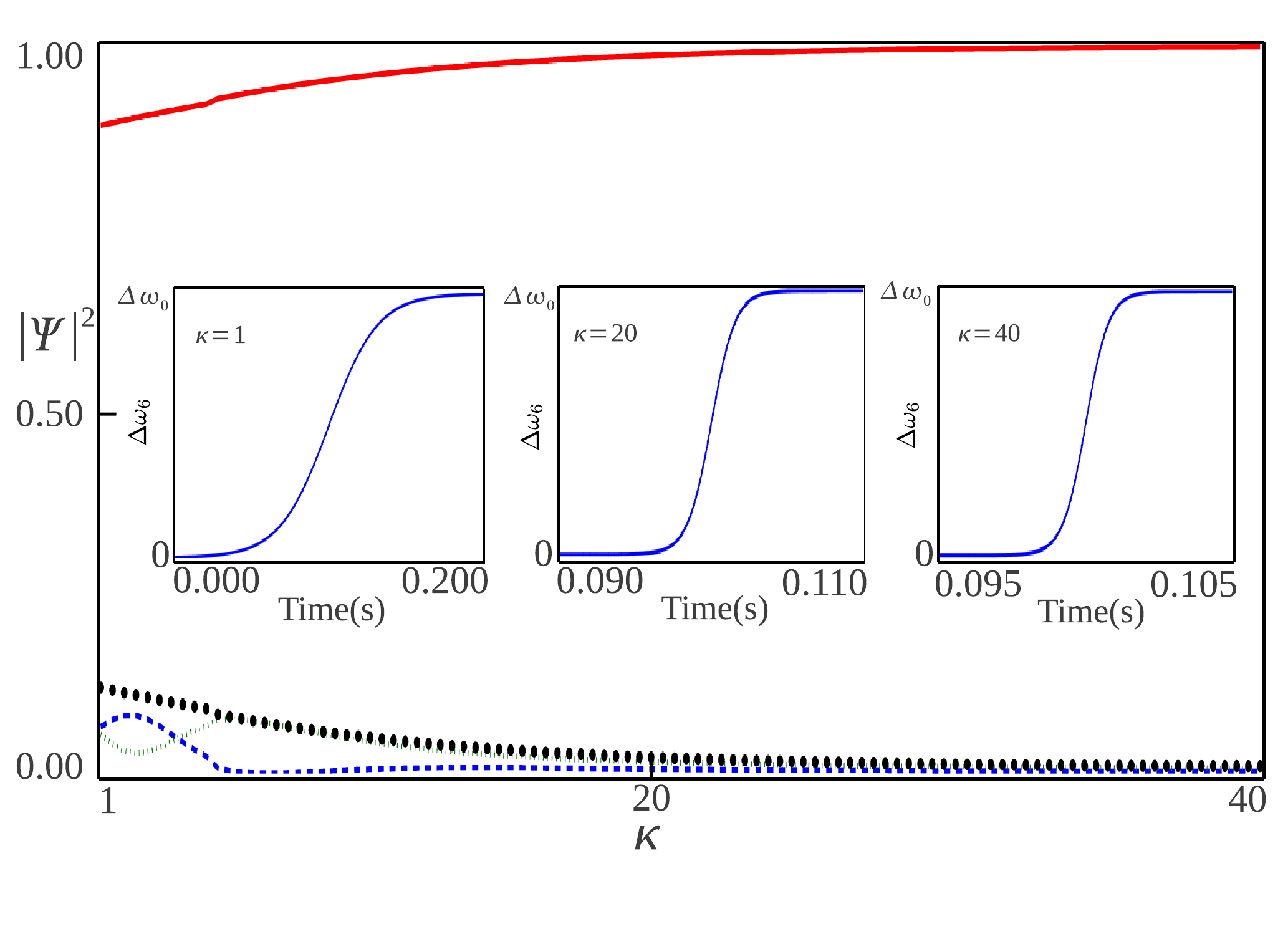}
  \caption{Final population in left (verticle dashed green line), middle
    (horizontal dashed blue line) and right (solid red line) traps for
    non-linear CTAP with increasing $\kappa$ and $\Delta \omega_0=1.5$ kHz. The
    dotted black line shows the total population not occupying the target
    (right) trap. The insets show the shape of $\Delta\omega_6(\kappa;\tilde t)$ for different 
    values of $\kappa$. An increased value of $\kappa$ increases the time 
    when the adjustment begins and decreases the adjustment time.}
\label{Fig:tanh}
\end{figure}

The outline of our scheme for dynamic detuning is as
follows. Initially the cloud is trapped in the left trap which we
detune such that resonance with the eigenstates of the other two traps
is ensured (since the traps are far apart, it is possible to determine
the chemical potential $\mu_L$).  As we time evolve the system
tunnelling sets in and we begin to reduce the detuning on the left
trap to zero while increasing the detuning of the right trap, as atoms
enter it. This can be achieved by adjusting the radio frequencies
$\omega_2$ and $\omega_6$, associated with the left and right hand
side trap, respectively. Here we suggest that a good form of function for the 
adjustment related to the left hand side trap is
\begin{equation}
  \Delta\omega_2(\kappa;\tilde t)=\frac{1}{2}[1-\tanh(\kappa
  \tilde t)]\Delta\omega_0 \;,
\end{equation}
where the inital value for the change in $\omega_2$ is given by
$\omega_0$. The function runs between $\Delta\omega_0$ and 0 and the
steepness in the crossover region is determined by $\kappa$. This
gives us an effective handle on both, the time when the adjustment
starts, and the duration of the adjustment (see inset of
Fig.~\ref{Fig:tanh}). Here $\tilde t=t-T/2$, with $T$ being the
overall duration of the process. At the same time the frequency of the
right hand side trap needs to be adjusted as well and it is easy to
see that a mirror symmetric change in $\omega_6$ is the best choice.
\begin{equation}
  \Delta\omega_6(\kappa;\tilde t)=\frac{1}{2}[1+\tanh(\kappa
  \tilde t)]\Delta\omega_0 \;,
\end{equation}
The dynamic adjustments of the radio frequency
equations~\eqref{eq:RFAdjustmentLinear} then become
\begin{subequations}
  \begin{align}
    \omega_1(t)&=\omega_1(t_0),\\
    \omega_2(t)&=\omega_2(t_0) - \Delta\omega_2(\kappa,\tilde t),\\
    \omega_3(t)&=\omega_3(t_0)-\frac{1}{2}f(t-\tau)\theta(t-\tau),\\
    \omega_4(t)&=\omega_4(t_0)-f(t-\tau)\theta(t-\tau),\\
    \omega_5(t)&=\omega_5(t_0)-\frac{1}{2}f(t)-f(t-\tau)\theta(t-\tau),\\
    \omega_6(t)&=\omega_6(t_0)-f(t) - f(t-\tau)\theta(t-\tau) +
    \Delta\omega_6(\kappa,\tilde t)\;.
  \end{align}
\label{eq:RFAdjustmentNonLinear}
\end{subequations}

In Fig.~\ref{Fig:tanh} we show the final population transferred to the
right trap for increasing values of $\kappa$ and for
$\Delta \omega_0=1.5$ kHz. We can see that the dynamic adjustment of the
detunings of the outer traps allows us to achieve population transfer
of $>99 \%$, up from $84 \%$. This is an improvement over both
standard CTAP and fixed detuning in the weak interaction regime and,
in fact, returns to the transfer efficiency of single particle CTAP.

\section{Conclusions}
We have shown that radio frequency traps can be used as microtraps for
processes in which an adjustable tunneling strength is
required. Neighbouring trapping potentials can be overlapped without
significantly changing the underlying energy level structure. This
property has allowed us to create a triple well radio frequency
potential in which coherent transport using adiabatic passage can be
demonstrated. For a single atom, it was shown that complete transfer
between the left and right traps by utilizing the dark state of the
system is possible, maintaining the advantages of an absence of Rabi
oscillations and robustness against variation in system parameters.

For a cloud of weakly interacting atoms we have demonstrated a
technique that significantly improves the efficiency of CTAP by
dynamically detuning the outer traps. Our suggested setup is close to
experimental realities, avoids the large overhead of other suggestions
and can easily be extended to other adiabatic techniques.

\section{Acknowledgements} 
The authors would like to thank Peter Kr\"uger and Thomas Fernholz for
valuable discussions.  This work was supported by Science Foundation
Ireland under project numbers 05/IN/I852 and 10/IN.1/I2979.
BOS. acknowledges support from IRCSET through the Embark Initiative
No. RS/2006/172.

\end{document}